# Genetic proof of chromatin diminution under mitotic agamospermy

Evgenii V. Levites

Institute of Cytology and Genetics, Siberian Branch of the Russian Academy of Sciences, Novosibirsk, Russia

Email: levites@bionet.nsc.ru

**ABSTRACT**

The previously published data are examined on the base of the hypothesis about existence of differential polyteny of chromosomes and excess chromatin diminution during the first stages of embryogenesis. It has been concluded that available data provide the genetic proof that chromatin diminution is one of the mechanisms underlying the origin of polymorphism in sugar beet agamospermous progenies.

A bulk of evidence has been obtained to date for polymorphism in agamospermous progenies of diploid plants. One of the explanations of such polymorphism is based on the recognition of the important role of mixoploidy in plants. Mixoploidy is manifested by an admixture of tetraploid cells among the bulk of diploid archespore cells of a mother plant (Maletskii, Maletskaya, 1996). Entering of a tetraploid cell into meiosis leads to the formation of diploid embryo sac and, accordingly, to the formation of diploid egg cell, capable of entering into embryogenesis without fertilization. This mechanism is characteristic of meiotic diplospory, which can be also designated as meiotic agamospermy (Levites, 2002). In this case polymorphism is a natural consequence of meiosis and can be designated by a known term "autosegregation" (Gustafsson, 1946-1947; Maletskii et al., 1998). Genetic and cytological data support this hypothesis (Shkutnik, 2010).

At the same time, an additional mechanism has been proposed to explain polymorphism in agamospermous progenies (Levites, 2005, 2007). It suggests that polymorphism occurs mostly due to the polytenization of marker loci bearing regions of chromosomes. Differential polyteny could subsequently lead to random equiprobable loss of excess chromatin by a cell before it enters embryogenesis.

Theoretical calculations indicate that the differential polytenization of chromosomes and subsequent diminution of excess chromatin is possible both under meiotic agamospermy and under mitotic agamospermy (adventive embryony) when the offspring arises from the somatic cells which have not undergone meiotic transformations of genome. There is also evidence that polytenization can occur in chromosome regions of egg cells under sexual plant reproduction (Levites and Kirikovich, 2013a). Genetic proof of this hypothesis has been obtained along with the proof that the process of polytenization depends on external conditions (Levites and Kirikovich, 2013b).

The studies of agamospermous progenies as well as the consideration of chromosome polytenization provide a new insight into many genetic processes and the causes of numerous variations in the genotype and phenotype ratios of the resulting offspring. A characteristic feature of polymorphism under agamospermy is the mismatch between the identified ratios and the normal Mendelian ratios.

Accounting for the effect of polytenization of chromosome regions bearing marker genes on the segregation of respective marker traits expands the boundaries of genetics. At present, trait segregation can be attributed both to changes in the number of chromosomes in the cell (meiosis and gamete fusion) and to other process not attributable to such changes (diminution of endoreduplicated sites of chromatids).

The facts collected since the early studies have contributed to a gradual shift in our view at the mechanisms underlying agamospermy. At this stage it is necessary to review our earlier data, which is the aim of this article.

Under discussion will be the data presented in the article entitled "Pseudosegregation in the agamospermic progeny of male sterile plants of the sugar beet (*Beta vulgaris* L.)" (1999), (Authors: Levites E.V., Shkutnik T., Ovechkina O.N. and Maletskii S.I.). In the paper the genetic methods were used to show that the analyzed sugar beet agamospermous progenies were formed from somatic cells. This conclusion was based on the monomorphism of the KWS1-5A offspring by heterozygous isozyme spectrum of marker enzyme alcohol dehydrogenase (ADH1). Interestingly, the study also revealed the dimorphism of the analyzed progenies, including KWS1-5A, for other marker enzymes. In the progeny the dimorphism of enzymes was expressed by the presence of only two phenotypic classes: one homozygous and one heterozygous. Of all the data from the cited article let us consider two offsprings, KWS1-5A and KHBC2-78A (Table 1).

Table1

Phenotypic classes of marker enzymes in agamospermous progenies of pollen sterile sugar beet plants

| Progeny | Marker enzyme phenotypes of progenies | | |
|---|---|---|---|
| | ADH1 | IDH3 | MDH1 |
| | FF : FS : SS | FF : FS : SS | FF : FS : SS |
| KWS1-5A | 0:78:0 | 23:41:0* | 9:0:0 |
| KHBC2-78A | - | 9:47:10** | 45:57:0* |

Probability of affinity with theoretically expected ratio 3:8:3 - *- $P<0.001$; ** - $P>0.05$

In the cited article it was assumed that the dimorphism was due to the inactivation in a part of the offspring of one of the alleles at a heterozygous locus. As a result, the seeds with the phenotypes similar to the homozygous one carry one active allele which determines the electrophoretic mobility of the enzyme and one inactivated allele. However, later it was found that phenotypes similar to those of the homozygous are conditioned by homozygous genotypes indeed (Levites, Kirikovich, 2003).

The findings allowed us to hypothesize that the dimorphism of agamospermous progeny is due to the heterozygosity at the marker enzyme locus with one allele represented by a single copy and the other allele represented by three copies arising as a result of polytenization (Levites, 2005, 2007). Polyteny of chromosomes in plants - a well known fact (Carvalheira, 2000). The somatic cells with the genotype *FFFS* lose excess copies of the alleles equiprobably before entering the embryogenesis. The calculations with the help of the hypergeometric distribution formulas indicate that in this case only two genotypes, *FF* and *FS*, in the ratio of 1:1 are theoretically possible.

The calculation is as follows:

For homozygotes FF - $C^2_3 \times C^0_1 / C^2_4$ - the number of combinations of choice two out of three, multiplied by the number of combinations 0 out of 1 and divided by the number of combinations 2 out of 4, i.e., 3x1/6.

For heterozygotes FS - $C^1_3 \times C^1_1 / C^2_4$ - the number of combinations of choice one out of three multiplied by the number of combinations one out of one, and divided by the number of combinations 2 out of 4, i.e., 3x1/6.

In the reduced form this ratio expressed in integers is 1:1.

The SS genotype is not formed because it requires two allele copies while only one copy is present in the genome.

Equiprobable diminution process requires free exchange of chromatides between chromosomes. Existence of such exchange was demonstrated later on the base of the phenotype ratio observed in the agamospermous progeny of a plant treated with colchicine (Levites, Kirikovich, 2012).

Therefore, it is interesting to consider the phenotype ratios of the marker enzymes isocitrate dehydrogenase (IDH3) and malate dehydrogenase (MDH1) in the agamospermous progeny KHBC2-78A. This offspring is of particular interest because it combines two distinctive traits that are inherent to agamospermous progeny: the phenotype class ratio for IDH3 corresponding to 3:8:3 and dimorphism for MDH1. The MDH1 phenotype class ratio indicates that this progeny has originated from somatic cells with different doses of alleles at *Mdh1* locus. The somatic origin of these cells implies that no meiotic genome transformations have occurred in the nuclei of such cells.

Mathematically, the ratio 3:8:3 is known to be possible derived if out of a sample containing 4 elements of one type and 4 elements of the other type 2 elements are selected randomly (Feller, 1950). This occurs, for example, when a heterozygous tetraploid cell of genotype *FFSS* enters into meiosis. Since at this moment each chromosome is represented by two chromatids, 8 copies of alleles are present in the nucleus by 4 copies of each of the two alleles.

If the frequency of crossing-over between the marker locus and the centromere is 50%, all copies of the alleles behave independently and the random selection of two copies obeys probability laws. The frequencies of the resulting gametes can be calculated by the hypergeometric distribution formulas (Feller, 1950). For the above example the gamete frequencies in fractions of units can be determined as follows:

For homozygotes *FF* - $C^2_4 \times C^0_4 / C^2_8$, the number of combinations of choice 2 out of 4 multiplied by the number of combinations 0 out of 4 and divided by the number of combinations 2 out of 8, i.e., 6x1/28.

For heterozygotes *FS* - $C^1_4 \times C^1_4 / C^2_8$, the number of combinations of choice 1 out of 4 multiplied by the number of combinations 1 out of 4 and divided by the number of combinations 2 out of 8, i.e., 4x4/28.

For homozygotes *SS* - $C^0_4 \times C^2_4 / C^2_8$, the number of combinations of choice 0 out of 4 multiplied by the number of combinations 2 out of 4 and divided by the number of combinations 2 out of 8, i.e., 6x1/28.

In the reduced form this ratio expressed in integers is 3:8:3.

From the above it can be concluded that the ratio 3:8:3 for IDH3 which observed in the KHBC2-78A offspring implies that: 1) the offspring emerged from the cells with an increased number of copies of each allele at the locus Idh3; 2) the number of copies of the alleles decreases in the moment before cells are entering into embryogenesis.

On the other hand, the presence in the same seeds of this progeny of two phenotypic classes for MDH1 indicates that this progeny originates from the cells which have not undergone meiotic genome transformations. Therefore, the reduction in the number of allelic copies at the Idh3 locus is not a consequence of meiosis but is the result of chromatin diminution only.

Moreover, the phenotype class ratios in both progenies described here can be explained precisely by chromatin diminution.

The presence in one offspring of two complementary traits (somatic origin of the cells entering into embryogenesis and an increased dose of alleles in the cells capable to embryogenesis) confirms both the agamospermous origin of the offspring and the process of chromatin diminution from the cells in the moment before their entering into embryogenesis.

In conclusion, it should also to add that according to the proposed hypothesis the equiprobable process of diminution of the number of redundant allelic copies is a consequence of equiprobable attachment of allelic copies to the nuclear membrane (Levites, 2005, 2007). It is assumed that only one copy from each chromosome out of two homologous in a diploid plant attaches to the nuclear membrane of the cell before its entering embryogenesis. Attached allelic pair determines the genotype of a developing embryo while the unattached allelic copies are lost.

Thus, a new analysis of the previously published data gives new insight on complementary genetic facts, which in aggregate confirm a model describing the specific mechanism of origin of polymorphism in agamospermous progenies. Available data provide the genetic proof that chromatin diminution is one of the mechanisms underlying the origin of polymorphism in sugar beet agamospermous progenies.